\documentclass[aps,floatfix,twocolumn,final,prl,showpacs]{revtex4}

\usepackage{epsfig}
\usepackage{graphicx}
\usepackage[ansinew]{inputenc}
\usepackage{array}
\usepackage{braket}
\usepackage{hyperref}
\usepackage{tabularx}
\newcolumntype{C}[1]{>{\centering\let\newline\\\arraybackslash\hspace{0pt}}m{#1}}

\usepackage{xcolor}
\usepackage{ulem}

\begin{document}

\preprint{APS/123-QED}
\title{Surface excitations relaxation in the Kondo insulator Sm$_{1-x}$Gd$_{x}$B$_{6}$}

\author{J. C. Souza$^{1,2}$, M. K\"onig$^{2}$, M. V. Ale Crivillero$^{2}$, M. O. Malcolms$^{1}$, R. R. Urbano$^{1}$, Z. Fisk$^{3}$, P. F. S. Rosa$^{4}$, P. G. Pagliuso$^{1}$, S. Wirth$^{2}$, and J. Sichelschmidt$^{2}$}

\affiliation{$^{1}$Instituto de F\'isica \lq\lq Gleb Wataghin\rq\rq,
UNICAMP, 13083-859, Campinas, SP, Brazil\\
$^{2}$Max Planck Institute for Chemical Physics of Solids, D-01187 Dresden, Germany\\
$^{3}$Department of Physics and Astronomy, University of California, Irvine, California 92697,USA\\
$^{4}$Los Alamos National Laboratory, Los Alamos, New Mexico 87545, USA}


\date{\today}

\begin{abstract}
The interplay between non-trivial topological states of matter and strong electronic correlations is one of the most compelling open questions in condensed matter physics. Due to experimental challenges, there is an increasing desire to find more microscopic techniques to complement the results of more traditional experiments.
In this work, we locally explore the Kondo insulator Sm$_{1-x}$Gd$_{x}$B$_{6}$ by means of electron spin resonance (ESR) of Gd$^{3+}$ ions at low temperatures. Our analysis reveals that the Gd$^{3+}$ ESR line shape shows an anomalous evolution as a function of temperature, wherein for highly dilute samples ($x$ $\approx$ 0.0002) the Gd$^{3+}$ ESR line shape changes from a localized ESR local moment character to a diffusive-like character. Upon manipulating the sample surface with a focused ion beam we demonstrate, in combination with electrical resistivity measurements, that the localized character of the Gd$^{3+}$ ESR line shape is recovered by increasing the penetration of the microwave in the sample. This provides compelling evidence for the contribution of surface or near-surface excitations to the relaxation mechanism in the Gd$^{3+}$ spin dynamics. Our work brings new insights into the importance of non-trivial surface excitations in ESR, opening new routes to be explored both theoretically and experimentally.
\end{abstract}

\maketitle

\section{\label{sec:intro}I. Introduction}

The concept of topology in condensed matter physics emerged from breakthroughs in the quantum Hall effect \cite{klitzing1980new}; however, more recently such concept has been generalized to other states of matter, such as topological insulators \cite{hasan2010colloquium}, Dirac and Weyl semimetals \cite{young2012dirac,yan2017topological,manna2018heusler}, and other exotic phenomena \cite{xu2012hedgehog,wang2016hourglass}. The gapless spin-polarized surface states of topological insulators were the first of these new states of matter to be explored, both theoretically and experimentally \cite{hasan2010colloquium}. One of the most pressing questions that remains open is the role of topological states of matter in systems where electronic correlations are important, which are known as strongly correlated systems \cite{dzero2012theory,alexandrov2013cubic,dzero2016topological,guo2018evidence,paschen2020quantum}.

The prototypical material to study this interplay is the Kondo insulator SmB$_{6}$ \cite{alexandrov2013cubic,rosa2020bulk2,li2020emergent}. Although extensively studied during the last 40 years, the prediction of a topological insulating ground state in this compound brought back interest in this system \cite{kim2014topological,kim2013surface,fuhrman2015interaction,neupane2013surface,jiang2013observation,suga2014spin,xu2014direct,wolgast2013low}. With a simple cubic structure (space group $Pm\overline{3}m$), SmB$_{6}$ has all the properties required of a cubic topological Kondo insulator (TKI), such as the $\Gamma_{8}$ quartet crystal field ground state and an odd number of band inversion at the $X$ point in the Brillouin zone \cite{li2020emergent,sundermann2018fourf,fuhrman2015interaction}. The hybridization between the $d$ conduction electrons and the Sm-4$f$ electrons opens a gap at the Fermi energy, and the system becomes a good insulator at low temperatures \cite{eo2018robustness,eo2020comprehensive}. In consequence, the bulk carriers contribute less to the transport upon lowering the temperature, and a plateau is observed in resistivity at low temperatures, which is related to surface states dominating the conductivity \cite{kim2014topological,eo2020comprehensive,zhang2013hybridization,syers2015tuning,lee2016observation}. An additional energy scale inside of the hybridization gap in scanning tunneling spectroscopy has been linked to these surface states \cite{jiao2016additional,jiao2018magnetic,pirie2018imaging}.

Although the surface states are well established, there is still debate about their topological nature. Although recent quasiparticle interference and angle-resolved photoemission spectroscopic results have demonstrated compelling evidence of non-trivial topology \cite{neupane2013surface,suga2014spin,xu2014direct,pirie2018imaging,li2020emergent,matt2020consistency}, other reports argue that a trivial surface state is at play \cite{hlawenka2018samarium,herrmann2020contrast}. Recent results show that the Kondo insulating state is very sensitive to disorder \cite{valentine2016breakdown,valentine2018effect,sen2018fragility,abele2020topological,souza2020metallic}, which indicate that the observed difference may be accounted for by subtleties in growth conditions \cite{rosa2020bulk2,hatnean2013large,li2014two,tan2015unconventional,phelan2016chemistry,thomas2018quantum,gheidi2019intrinsic}. Although natural impurities, such as Gd$^{3+}$, locally affect the hybridization gap, highly dilute concentrations do not globally affect the Kondo insulating phase \cite{souza2020metallic,fuhrman2018screened}. In particular, Sm$_{1-x}$Gd$_{x}$B$_{6}$ samples with $x$ = 0.0002 exhibit insulating behavior at low temperatures \cite{crivillero2021resistivity}. In this scenario, the use of a spectroscopic technique to complement the recent experimental results is highly desirable. Although electron spin resonance (ESR) is a bulk sensitive measurement, the microwave penetration into the sample can be affected by both surface and bulk conductivity. Furthermore, the resonance energy absorbed by the probe spin may ultimately relax through the surface to the thermal bath, which makes ESR also surface sensitive \cite{feher1955electron,lesseux2016unusual,demishev2018magnetic}. 

\begin{figure*}[!ht]
\includegraphics[width=0.99\textwidth]{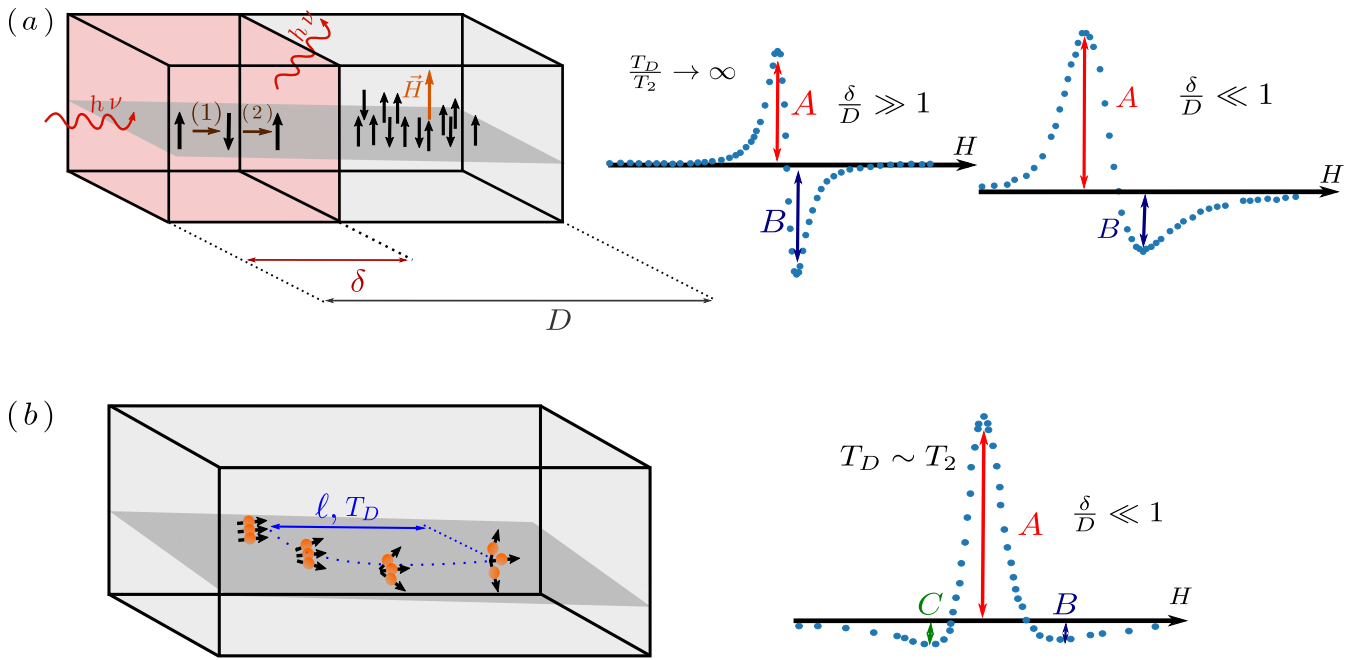}
\caption{(a) Pictorial representation of a local moment ESR of a $S$ = 1/2 probe, where $\delta$ is the skin depth, $D$ the thickness of the sample, $T_{D}$ the diffusion time and $T_{2}$ the spin-spin relaxation time. $\uparrow$ represents the $\ket{\uparrow}$ ground state and $\downarrow$ the $\ket{\downarrow}$ excited state. The orange vector $\mathbf{H}$ shows the externally applied magnetic field direction. $\delta/D$ $\gg$ 1 is an insulator (symmetric \lq\lq local\rq\rq line shape) and $\delta/D$ $\ll$ 1 is a metal - \lq\lq local\rq\rq line shape with \lq\lq skin asymmetry\rq\rq. (b) Pictorial semi-classical representation of the interplay between the coherence loss of the spin system ($T_{2}$) and the diffusion of the spin system ($T_{D}$, which is connected with the mean free path $\ell$) for a sample showing a CESR. The line shape represents the CESR case when $T_{D}$ $\sim$ $T_{2}$ (line shape with \lq\lq diffusive asymmetry\rq\rq). The dotted line in the left panel illustrates the spatial shift of the carrier. For simplicity we do not present the concomitant spin-lattice relaxation of the spin system. The ESR spectra in (a) and (b) are plotted as the power absorption derivative ($dP/dH$) as a function of the magnetic field $H$ and were adapted from \cite{hemmida2018weak}, \cite{pagliuso1999electron}, \cite{walmsley1988spin} for the insulator, metal and conduction-electron case, respectively.}
\label{Fig01}
\end{figure*}

The relation between ESR properties and sample conductivity has been investigated for the case in which diffusion of spin excitations becomes relevant \cite{barnes1981theory}. F.J. Dyson has demonstrated early on \cite{dyson1955electron} that the line shape can basically be influenced by two effects: the skin effect, through the ratio of the skin depth $\delta$ to the sample thickness $D$, and the diffusion of the resonating magnetic moments, through the ratio between diffusion time $T_{D}$ and the spin-spin relaxation time $T_{2}$. On one hand, the former is a result of the interplay of the microwave field and the carriers, where shielding currents drive electromagnetic fields out of phase \cite{barnes1981theory,poole1971relaxation,abragam2012electron,hemmida2018weak}. On the other hand, the latter is a measure of the coherence loss of the resonating spins \cite{barnes1981theory,poole1971relaxation,abragam2012electron}. 

The case for a local moment ESR is pictorially represented in figure \ref{Fig01} a) for a simple $S$ = 1/2 system.
An external magnetic field splits the degenerated spin state into a spin-up ground state $\ket{\uparrow}$ and a spin-down excited state $\ket{\downarrow}$. Microwave energy matching this energy splitting can be absorbed [process (1)] and emitted, process (2). A net absorption, and therefore an observable ESR signal is possible when the spin system relaxes energy from $\ket{\downarrow}$ to the thermal bath. Here, two different and simultaneous relaxation mechanisms are involved: the already mentioned spin-spin relaxation, and the spin-lattice relaxation, which is connected to the energy transfer through phonons to the thermal bath with a characteristic time scale $T_{1}$ \cite{orbach1961theory,garanin2015angular,nakane2018angular}. 

The skin effect and the diffusion of spin excitations lead to distortions of the line shape which can be measured by the amplitude ratio $A/B$ as shown in fig. \ref{Fig01} - the ESR spectra were adapted from \cite{hemmida2018weak,pagliuso1999electron,walmsley1988spin}. In the almost completely stationary regime of local moment ESR, $i.e.$ $T_{2}$ $\ll$ $T_{D}$, the symmetry of the line shape is only defined by the ratio $\lambda = \delta/D$ \cite{feher1955electron}. As shown in the center panel of Fig. \ref{Fig01} a) for insulators, where $\lambda\gg1$, one obtains a symmetric Lorentzian line shape with $A/B=1$. In metals, where $\lambda\ll 1$, the skin depth effect will be more relevant and a so-called Dysonian line shape occurs, which has a \lq\lq skin asymmetry\rq\rq and an upper limit $A/B\approx2.7$ \cite{feher1955electron,barnes1981theory}. The differences between insulators and metals are further manifested in the relaxation of the system - which can be explored by measuring the saturation of the ESR intensity as a function of the microwave power, where insulators saturate at much lower microwave powers when compared with conductors \cite{poole1971relaxation,supp2020souza}.

Fig. \ref{Fig01} b) depicts in a semi-classical way the case for a conduction-electron spin resonance (CESR), where $T_{D}$ can be comparable in some cases to $T_{2}$. 
The interplay between the coherence loss and the diffusion of the spin system plays an important role on the ESR line shape. While the coherence loss, as governed by $T_{2}$, macroscopically results in the loss of the transverse magnetization, the diffusion, characterized by $T_{D}$, is connected with the mean free path $\ell$. A line shape with a \lq\lq diffusive asymmetry\rq\rq is obtained for $T_{D}$ $\sim$ $T_{2}$, resulting in $A/B$ $\geq$ 2.7 and an additional $C$ valley in the shape, as defined in fig. \ref{Fig01} b). If $T_{D} \sim T_{2}$, the spin probe has a significant probability of a considerable spatial shift before the coherence between the probe spins is lost. Therefore, there is a diffusion of the spin excitations within the layer induced by the skin effect, which results in a diffusive line shape, shown in the right panel of fig. \ref{Fig01} b). The spin-lattice relaxation, which macroscopically results in the recovery of the longitudinal magnetization, is not represented in fig. \ref{Fig01} b) for simplicity. In systems with a mean free path greater than the skin depth, $i.e.$ $\ell$/$\delta$ $\geq$ 1 (anomalous skin effect), the ratio $C/B$ can be larger than 1 \cite{pifer1971conduction}.

Recently, highly unusual ESR line shapes were reported for the half-Heusler compounds YPtBi and YPdBi substituted with Nd$^{3+}$ \cite{lesseux2016unusual,souza2018diffusive}. The Nd$^{3+}$ ESR line shape showed a diffusive asymmetry although the Nd$^{3+}$ spins are localized. This diffusive asymmetry was discussed to be an experimental signature of nontrivial topological states. It was related to a relaxation mechanism through Dirac excitations in or near the surface by virtue of a so-called phonon-bottleneck effect, which results in an enhanced spin-lattice relaxation time $T_{1}$ \cite{orbach1961theory,garanin2015angular,nakane2018angular,garanin2007towards}. In this case, the absorbed energy would diffuse through the surface before relaxing to the thermal bath. 

In this work, we report the observation of a diffusive asymmetry in the Gd$^{3+}$ ESR line shape of highly dilute Gd$^{3+}$-substituted SmB$_{6}$. Combining microwave power-dependent ESR, focused ion beam (FIB) for cutting trenches on the sample surface, and complementary resistivity measurements, we provide evidence for surface excitations contributing to the ESR relaxation. The FIB treatment of the sample surface results in an increase of the skin depth \cite{supp2020souza}, $i.e.$ the microwave penetration, and the recovery of a local ESR line shape \cite{lesseux2016unusual}. Our temperature and FIB dependencies of the diffusive-like line shape and the Gd$^{3+}$ ESR spin relaxation provides strong evidence that surface and near-surface excitations, in the presence of a phonon bottleneck regime, are crucial ingredients to obtain such an unusual effect in an ESR experiment. As this system has been claimed to have metallic surface states \cite{neupane2013surface,xu2014direct,pirie2018imaging,li2020emergent,matt2020consistency}, these surface excitations are likely to be electronic.

\section{\label{sec:experiment}II. Methods}

Single crystalline samples of Sm$_{1-x}$Gd$_{x}$B$_{6}$ ($x$ = 0.0004 and 0.0002) were synthesized by the Al-flux growth technique with starting elements Sm:Gd:B:Al in the proportion of (1 - $x$):$x$:6:600 \cite{rosa2020bulk,rosa2020bulk2,thomas2018quantum}. The samples ranged in size from $\sim$ 0.7 to 1.4 mm width, 300 to 900 $\mu$m length and 120 to 500 $\mu$m thickness. Laue measurements confirmed the (001) planes of the largest facets. The $x$ used in the text refers to the nominal Gd$^{3+}$-concentration value. The magnetic properties of $x$ = 0.0002 samples were obtained using a vibrating sample magnetometer equipped with a superconducting quantum interference device (SQUID-VSM) \cite{supp2020souza}. Electrical resistivity was measured using a four-point technique with van der Pauw-geometry.

The crystals were etched before the first ESR measurement in a dilute mixture of hydrochloric and nitric acids in a proportion of 3:1 to remove possible impurities on the surface of the crystals due to Al flux. We did not polish any of the crystals in this study. For $x$ = 0.0002 we show the results of two different crystals (S1 and S2). The ESR measurements were performed on single crystals in a X-band ($\nu$ $\cong$ 9.4 GHz) spectrometer equipped with a goniometer and a He-flow cryostat in the temperature range of 2.6 K $\leq$ $T$ $\leq$ 40 K at powers of 0.2 $\mu$W $\leq$ $P$ $\leq$ 10 mW. The ESR spectra were analyzed using the software Spektrolyst.

In order to investigate the dependence of the ESR line shape on the surface properties, we employed a focused ion beam (FIB) for surface treatment using a Xe ion beam with currents of 500 nA and acceleration voltage of 30 kV. It turned out that this technique can change the surface conductivity in a systematic way, in contrast to just using a hand-made surface scratch \cite{crivillero2021resistivity}. In each FIB treatment we cut linear trenches of about 7-10 $\mu$m depth into the sample surface, resulting in a grid of such trenches with ever-increasing density. In the first two treatments we divided the sample in four equal parts (F1 and F2). In subsequent runs we approximately doubled the number of lines in each direction \cite{crivillero2021resistivity}. We investigated the ESR after each new FIB cut. We also performed energy dispersive x-ray spectroscopy in regions which were milled a few $\mu$m into the bulk. It was possible to only detect the signals of Sm, B and O, with Gd being below the detection limit (usually $\sim$ 1 \%) \cite{crivillero2021resistivity}. Interestingly, after milling no Al signal could be detected, indicating that the Al content in the bulk, if any, is below the detection limit - which differs from, $e.g.$, UBe$_{13}$ \cite{amon2018tracking}.

\section{\label{sec:resultsanddiscussion}III. Results}

\begin{figure}[!ht]
\includegraphics[width=0.99\columnwidth]{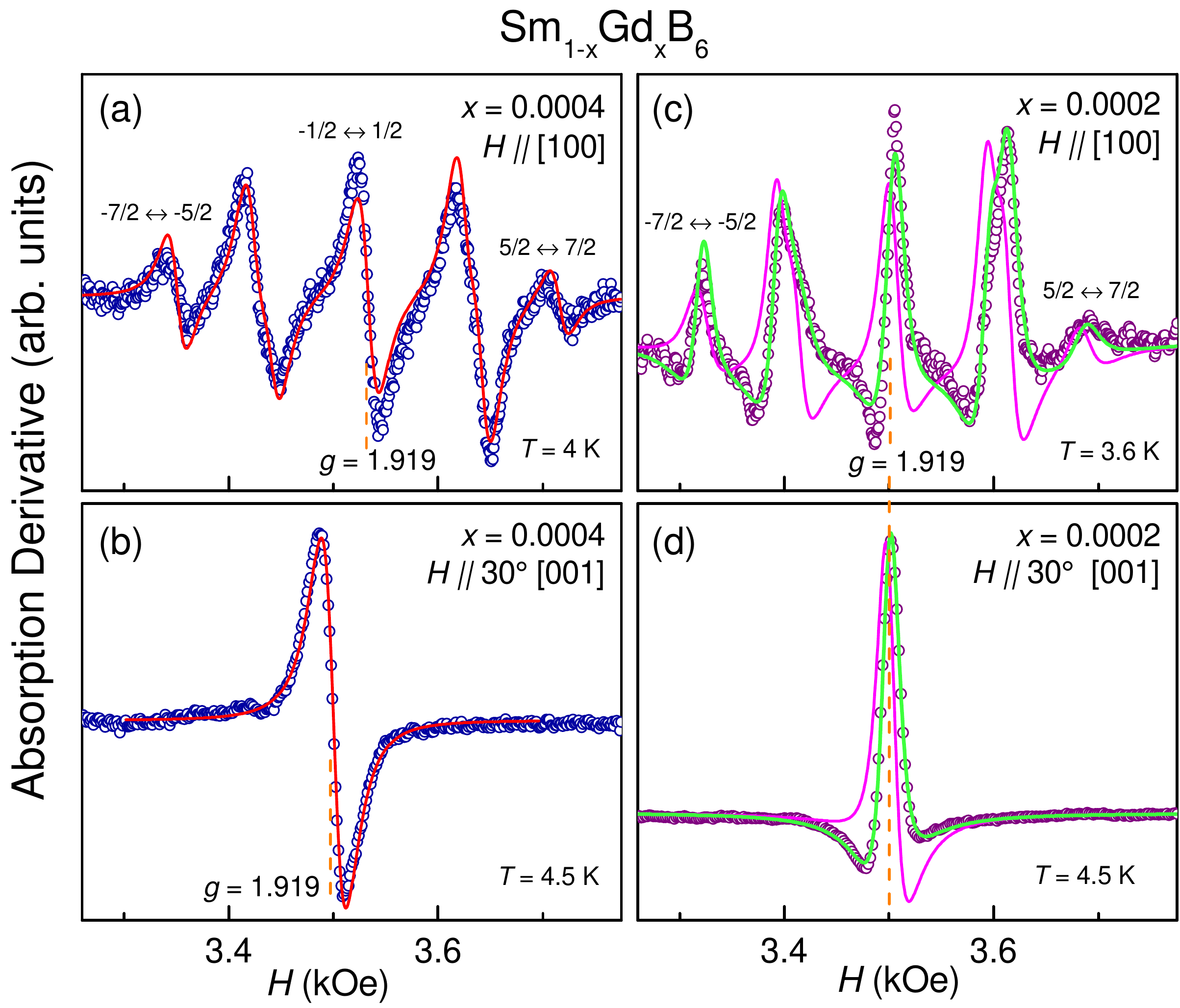} \caption{Gd$^{3+}$ ESR spectrum for $H$ applied parallel to the [100] direction and the Gd$^{3+}$ collapsed spectrum for (a), (b) $x = 0.0004$ and (c), (d) $x = 0.0002$ in Sm$_{1-x}$Gd$_{x}$B$_{6}$. The red, magenta and green solid lines are explained in the text. Two pairs of transitions ($\pm$ 5/2 $\rightarrow$ $\pm$ 3/2 and $\pm$ 3/2 $\rightarrow$ $\pm$ 1/2) in (a) and (c) are close in energy, which results in a weak shoulder in the left and right middle lines. The orange dashed lines show the $g$-value = 1.919, which is characteristic of Gd$^{3+}$ ions highly diluted inside a SmB$_{6}$ matrix \cite{souza2020metallic}. Gd$^{3+}$ ESR spectra for $x$ = 0.0004 have been adapted from \cite{souza2020metallic}.}
\label{Fig2}
\end{figure}

Figure \ref{Fig2} (a) shows a fine-structure split Gd$^{3+}$ ESR spectrum at $T$ = 4 K for Sm$_{1-x}$Gd$_{x}$B$_{6}$ ($x$ = 0.0004) with applied magnetic field $H$ parallel to the [100] direction. The well-resolved fine structure is characteristic of spin probes immersed in the insulating sample bulk allowing the weak crystalline electric field (CEF) of the Gd$^{3+}$ ions to split the line \cite{abragam2012electron,barnes1981theory}. Gd$^{3+}$ substitute Sm ions, which have a cubic local symmetry. As such, a cubic CEF effect is expected for Gd$^{3+}$ ions. In fact, the red solid line is a simulation with seven resonances considering a cubic CEF spin Hamiltonian with a Gd$^{3+}$ crystal field parameter $b_{4}$ = -9.5(3) Oe \cite{souza2020metallic}. Two pairs of fine-structure transitions are close in energy, which results in weak shoulders. In order to analyze our ESR line without the fine-structure influence, we turned the sample by 30 degrees away from the [001] towards the [110] direction until the Gd$^{3+}$ ESR spectrum is collapsed into one Gd$^{3+}$ resonance line \cite{barnes1981theory,abragam2012electron} as shown in Figure \ref{Fig2} (b). The red solid line is the best fit with a Lorentzian line shape, which is expected for an insulator.

Figure \ref{Fig2} (c) shows the Gd$^{3+}$ ESR spectrum for $x$ = 0.0002 at $T$ = 3.6 K for $H$ parallel to the [100] direction (sample S1). Again we observe seven resonances, which reinforces the notion that we are probing the bulk of SmB$_6$. However, we cannot reproduce our spectrum by using a local-type line shape with skin asymmetry (\lq\lq Dysonian\rq\rq) - see the magenta solid line. The difficulties in adjusting the data with a fine-split Dysonian line shape becomes even clearer when we collapse the spectrum into one line, as shown by the magenta solid line in Fig. \ref{Fig2} (d). However, as demonstrated by the green lines, a diffusive asymmetry in the line shape ($C$-valley $>0$) of the Gd$^{3+}$ ESR describes the spectra very well. This is highly unusual for a Gd$^{3+}$ spin probe that is expected to be localized in SmB$_{6}$. 

In order to quantify the development of this diffusive asymmetry, we propose an analogy to a model of a CESR in the presence of an anomalous skin effect \cite{pifer1971conduction}. In this model, the ESR spectrum, which is expressed by the power absorption derivative ($dP/dH$) as a function of $H$, can be described as

\begin{equation}
\frac{dP}{dH} \propto [1-(X/R)^{2}] \frac{d}{dx}\left ( \frac{1}{1+x^{2}} \right ) + (2X/R) \frac{d}{dx} \left ( \frac{x}{1+x^{2}} \right ),
\label{Eq1}
\end{equation}
where $x$ = $2(H - H_{r})/\Delta H$, with $H_{r}$ as the resonance field and $\Delta H$ as the line width \cite{feher1955electron}. The $X/R$ parameter in this CESR model is directly connected with the surface impedance, where $R$ is the surface resistance and $X$ the surface reactance \cite{pifer1971conduction,reuter1948theory}. For $X/R$ = 0 we have a symmetric Lorentzian line shape (A/B = 1), whereas $X/R=\sqrt2-1$ corresponds to a Dysonian line shape (A/B $\approx$ 2.7). Finally, $X/R$ $\geq$ 1 occurs when the anomalous skin effects play a role ($C/B$ $\geq$ 1) \cite{pifer1971conduction}.

The green solid line in fig. \ref{Fig2} (c) is a simulation assuming the same parameters for the $g$-values and $b_{4}$ from the red solid line in fig. \ref{Fig2} (a) and using $X/R$ = 1.6. For the Gd$^{3+}$ collapsed spectrum, shown in fig. \ref{Fig2} (d), we exclude the influence of the crystal field, therefore we can fit our data using eq. \ref{Eq1}. Again we maintained the same $g$-value and $\Delta H$ of the red solid line and obtained a $X/R$ $\approx$ 1.3. The simulation and the fit reproduce nicely the unusual spectra shape, which in case of a CESR would indicate that spin diffusion is relevant in the relaxation process. However, we recall that in Sm$_{1-x}$Gd$_{x}$B$_6$ ESR does not probe conduction electrons and, therefore, $X/R$ here is just a mathematical parameter. It is important to note that in fig. \ref{Fig2} (c) a significant diffusive asymmetry is present in each of the fine-split Gd$^{3+}$ lines, confirming again their origin from Gd$^{3+}$ ions in the bulk of SmB$_{6}$.

One possible explanation for the diffusive asymmetry could rely on an electrodynamics effect leading to an unconventional mixing of absorption and dispersion parts of the Lorentzian line shape in eq. (\ref{Eq1}). Such explanation could be based on highly conductive surfaces on top of an insulating bulk causing a large phase shift in the microwave response, making the dispersion to dominate. However, such interpretation, which is not consistent with Dyson's theory \cite{dyson1955electron,feher1955electron}, has not been observed experimentally and is not supported by at least two other previous experimental observations. The first, and most important, is the coexistence of localized and diffusive ESR line shapes in half-Heusler systems \cite{lesseux2016unusual,liu2016observation,souza2018diffusive}. Another important example is the local-moment ESR spectra in superconductors, which have highly conducting surfaces and do not show a diffusive-like character \cite{rettori1973magnetic,engel1973local,alekseevskii1973electron,davidov1974electron,orbach1974electron}. Instead, as expected from Dyson's theory, the maximum $A/B$ ratio is $\approx$ 2.7 \cite{feher1955electron,dyson1955electron,barnes1981theory}. Both examples support the notion that in Sm$_{1-x}$Gd$_{x}$B$_{6}$ the strong and peculiar asymmetry of the Gd$^{3+}$ line shape does not arise from an electrodynamics effect of the inhomogeneous conductivity cross section of the sample.

\begin{figure}[!ht]
\includegraphics[width=0.99\columnwidth]{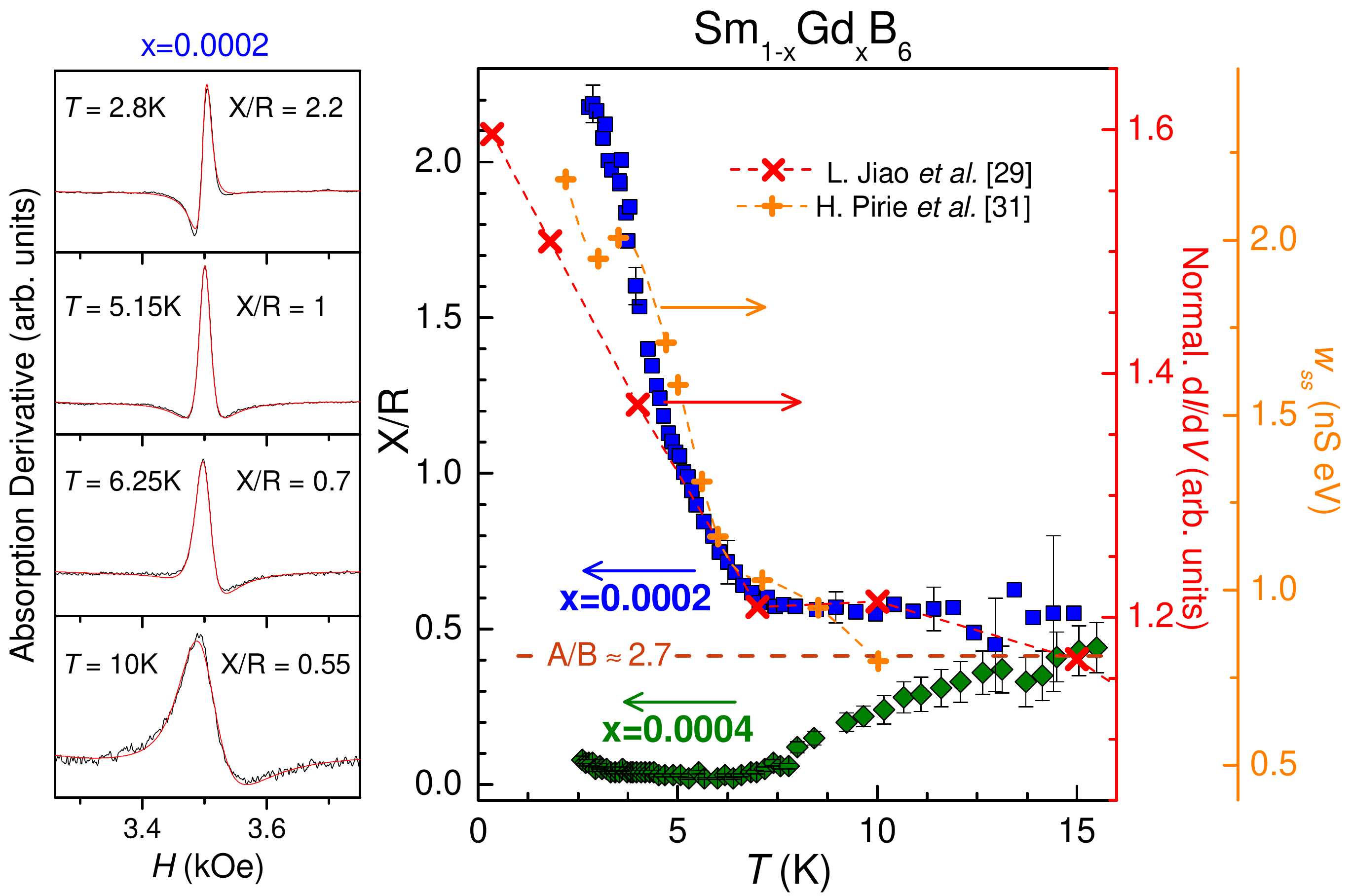}
\caption{Temperature dependence of Gd$^{3+}$ ESR spectra and line shape parameter $X/R$ in Sm$_{1-x}$Gd$_{x}$B$_6$. All the Gd$^{3+}$ ESR spectra were taken with a microwave power smaller than the saturation of the system. Left panel: ESR spectra (black lines) for $x$ = 0.0002 at various temperatures and spectra fits (red lines) by eq. (\ref{Eq1}) with $X/R$ parameters as indicated. Right panel: temperature dependence of $X/R$ for $x$ = 0.0002 and $x=0.0004$. A localized character of the line shape occurs below the brown dashed line which indicates a line shape asymmetry of $A/B\approx 2.7$. The red (orange) symbols show the $V_{b}$ = - 6.5 meV (- 5 meV) scanning tunneling spectroscopy peak intensity obtained in a non-reconstructed (reconstructed) B- (Sm-) terminated surface for SmB$_{6}$ reported previously \cite{jiao2016additional} (\cite{pirie2018imaging}). The orange and red dashed lines are just guides to the eyes.}
\label{Fig3}
\end{figure}

The right panel of figure \ref{Fig3} shows the temperature evolution of the collapsed Gd$^{3+}$ ESR spectra diffusive parameter $X/R$ for Sm$_{1-x}$Gd$_{x}$B$_{6}$ with $x$ = 0.0002 (sample S1) and $x$ = 0.0004. A similar evolution has been found for each line of the fine-split spectra at $H \|[100]$ \cite{supp2020souza}.
At temperatures exceeding $T\approx 12$~K, where the transport is dominated by carriers in the bulk of the sample \cite{jiao2016additional,eo2018robustness,souza2020metallic}, both samples show $X/R$ $\sim 0.5$, which indicates a localized-like behavior of the Gd$^{3+}$ ESR spectra, $i.e.$ a line shape with a skin asymmetry, $A/B\approx2.7$. For $x$ = 0.0004, the expected development towards a symmetric Lorentzian line shape ($X/R\sim0$) is observed at low temperatures, which is typical due to the insulating nature of the bulk. 
For $x = 0.0002$, at these temperatures, we should expect for $X/R$ a similar value, or at most $X/R\sim 0.5$ if the sample size is larger compared to the skin depth.
However, for $x = 0.0002$ the $X/R$ parameter strongly increases below $T\approx6$~K. This evolution is also clearly visible in the spectra as shown in the left panel of figure \ref{Fig3}. The decrease of the growing rate of the $X/R$ parameter towards the lowest temperatures should be taken with care due to the increase of error bars for higher $X/R$ values. \cite{supp2020souza}.

The $X/R$ temperature evolution is reminiscent of a signature peak in scanning tunneling spectroscopy studies. For comparison, the temperature dependencies of the intensity of the $V_{b}$ = - 6.5 meV and - 5 meV peaks are also included in the right panel of figure \ref{Fig3} (red and orange data points) \cite{jiao2016additional,pirie2018imaging}. The peaks, measured in a non-reconstructed B-terminated (- 6.5 meV) and in a reconstructed Sm-terminated SmB$_{6}$ surfaces (- 5 meV), were correlated to the surface states in pristine Al-flux grown samples \cite{jiao2016additional,pirie2018imaging}. As argued by Jiao $et$ $al.$, the clear change of the $T$-dependence of the intensity of the - 6.5 meV peak relies on the formation of the metallic surface states \cite{jiao2016additional}. 

The correlation between the evolution of both peaks, from differently terminated surfaces, and the evolution of a diffusive asymmetry of the line shape hints to the relevance of the surface states of SmB$_{6}$ in the Gd$^{3+}$ ESR line shape. In this respect, we should expect a relaxation mechanism including a coupling between the bulk Gd$^{3+}$ impurities and the surface states. Such coupling should be mediated by a relaxation through the phonons, which causes an enhancement of the spin-lattice relaxation time $T_{1}$ \cite{lesseux2016unusual}. Therefore, it would be helpful to tune the diffusive asymmetry of the Gd$^{3+}$ ESR line shape to construct an appropriate relaxation scenario. To this end, we changed the surface properties using a focused ion beam. A detailed resistivity study of the FIB effects can be found in ref. \cite{crivillero2021resistivity}. Another important tuning parameter is the Gd$^{3+}$ concentration, which can be a source to understand the role of disorder in the diffusive asymmetry of the line shape \cite{souza2020metallic}.

\begin{figure*}[!ht]
\includegraphics[width=0.75\textwidth]{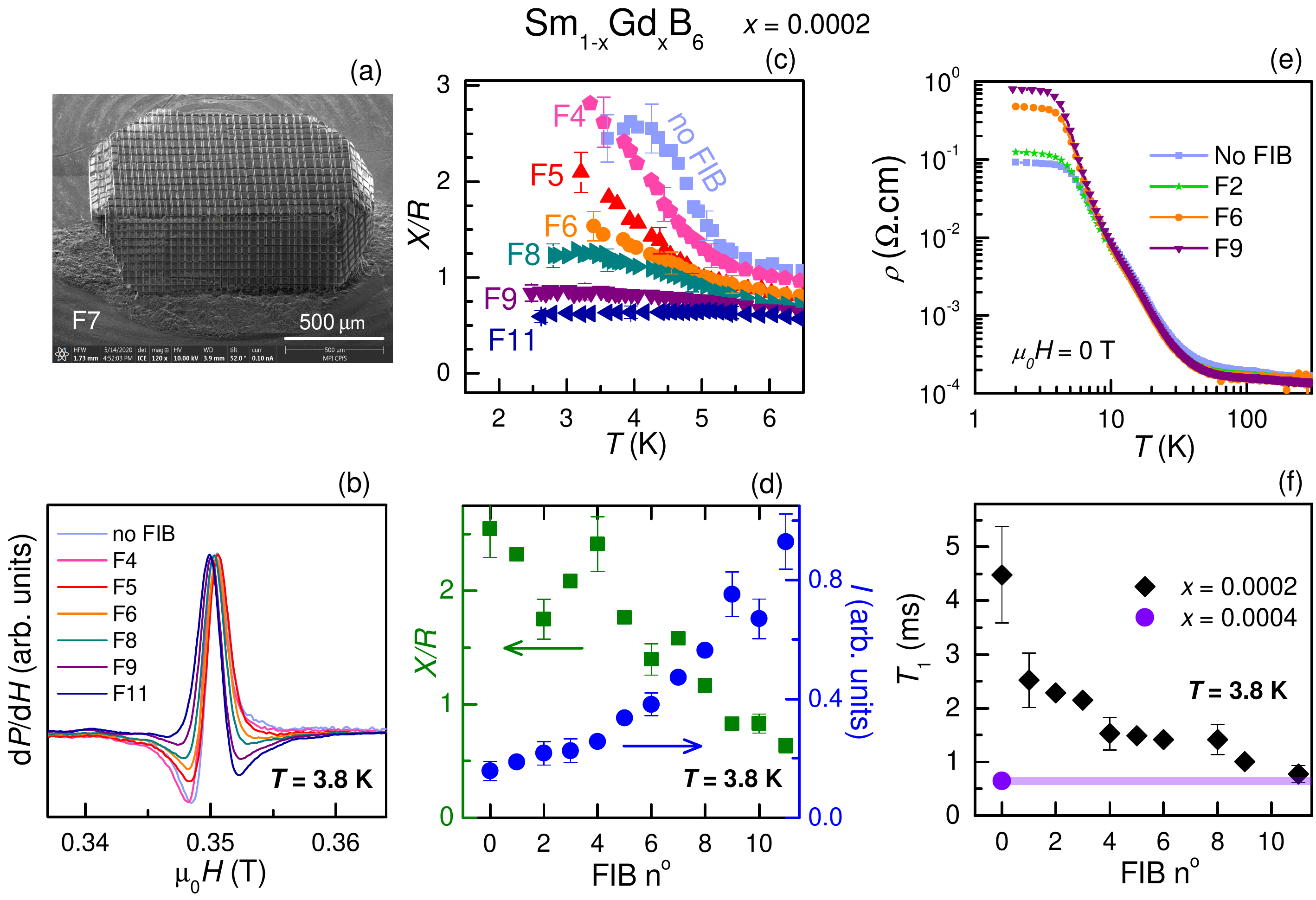}
\caption{(a) Scanning electron microscopy image of the Sm$_{0.9998}$Gd$_{0.0002}$B$_{6}$ sample S2 with intentional cuts made using a FIB. (b) Gd$^{3+}$ ESR spectra at $T$ $\approx$ 3.8 K as a function of the FIB grid. The spectra were normalized in order to make the ESR line shapes comparable. (c) $X/R$ line shape parameter as a function of temperature for different FIB cutting-grids. (d) FIB cutting-grid dependence of $X/R$ parameter and Gd$^{3+}$ ESR intensity at $T=3.8$~K. (e) Resistivity of sample S2 as a function of FIB cutting-grid. Figure adapted from \cite{crivillero2021resistivity}. (f) Spin-lattice relaxation time $T_{1}$ as a function of FIB cutting for $x$ = 0.0002 and 0.0004.}
\label{Fig4}
\end{figure*}

Figure \ref{Fig4} (a) exemplifies one of the stages of the FIB cuts in sample S2. As shown in figure \ref{Fig4} (b), at $T = 3.8$~K, there is a systematic change of the Gd$^{3+}$ diffusive-like line shape as a function of the FIB grid, which finally results in a Gd$^{3+}$ localized-like (skin asymmetric) line shape after the final removal (F11), in which no distinct trenches were cut but rather an approximate 5 $\mu$m thick layer was FIB-sputtered from the complete surface. We must notice that, as shown in table SI, the $g$-value and the Gd$^{3+}$ linewidth remain unchanged within our experimental uncertainty as a function of the FIB grid.

Figure \ref{Fig4} (c) presents the temperature evolution of the $X/R$ parameter for $T\leq 6.5$~ K and for selected FIB grids for Sm$_{1-x}$Gd$_{x}$B$_{6}$ ($x$ = 0.0002). At $T\geq5$~K for the first few FIB grids it was possible to observe $X/R>0.5$, $i.e.$ a diffusive asymmetry of the line shape. With increasing FIB grid, $X/R$ is systematically reduced. As shown in fig. \ref{Fig3}, the scanning tunneling peaks associated with the surface states are still present at these temperatures \cite{jiao2016additional,pirie2018imaging}. As such, we can expect that surface states may still play a role at these temperatures.
For $T$ $\leq$ 5 K we can see a systematic drop of the $X/R$ parameter as a function of the FIB-cutting. The evolution of the Gd$^{3+}$ diffusive asymmetry is heavily suppressed as a function of the FIB-cutting grid.

Figure \ref{Fig4} (d) shows the Gd$^{3+}$ ESR intensity \cite{supp2020souza} as a function of the FIB grid. This intensity is not only proportional to the Gd$^{3+}$ concentration but also to the interactive volume $V^{int}$ which, in a first approximation, is given by $V^{int}$ = $\sigma$$\delta$, with $\sigma$ being the surface area of the crystal. We further assumed that the FIB cutting has a small effect on the total volume of the sample, which can be included in the error bar.
As shown in figure \ref{Fig4} (d), there is a systematic increase of the Gd$^{3+}$ ESR intensity and, hence, the interactive volume as a function of the FIB grid. In other words, the skin depth gets larger as a function of increasing number of FIB-cut lines. At the same time, the $X/R$-development indicates an evolution from a diffusive to a skin asymmetry of the line shape. This is an important hint that surface effects should be considered as an essential ingredient to the unusual diffusive asymmetry of the line shape. The increase of the skin depth suggests a FIB-induced depletion of the surface states, which effectively increases the surface resistivity \cite{crivillero2021resistivity}. Such increase may be related with confinement of surface states or even disorder \cite{sen2018fragility,abele2020topological,sacksteder2015modification}. Disorder may also affect, for example, the Sm valence near the surface \cite{zabolotnyy2018chemical,fuhrman2019magnetic}.

The presence of conducting surface states has been demonstrated to be the origin of the resistivity plateau in SmB$_6$ \cite{kim2013surface,wolgast2013low,zhang2013hybridization}. Therefore, as a matter of comparison, we also measured the dc resistivity $\rho$ of the same ESR-investigated samples as a function of temperature for different FIB cutting-grids as shown in figure \ref{Fig4} (e) \cite{crivillero2021resistivity}. The value of the low-temperature resistivity plateau continuously increases with the grid of the FIB cuttings. Hence, in the low-temperature regime, for a FIB treated sample surface the contribution of the surface states to the overall conduction appears to be suppressed. Accordingly, the skin depth as determined from the resistivity should be affected as well.
Using a two layered model, where we consider that the conducting carriers at the surface are the main contributors to the resistivity at low temperatures \cite{zhang2013hybridization,syers2015tuning,supp2020souza}, we show that there is a systematic increase of the skin depth and a decrease of the mean free path $\ell$ as a function of the density of FIB-cut trenches - table SI \cite{supp2020souza}. 
Therefore, the increase of the skin depth estimated by dc resistivity and Gd$^{3+}$ ESR intensity (fig. \ref{Fig4} (d)) are consistent. However, when comparing skin depth results from both methods one must bear in mind that for ESR the local resistivity is the only relevant factor, while a non-local (global) character prevails in resistivity. It is therefore more sensitive to extrinsic effects, such as subsurface cracks, dislocations and any residual flux \cite{eo2018robustness,eo2020comprehensive,eo2020bulk,crivillero2021phase,crivillero2021resistivity} while local ESR measurements are not strongly influenced by these extrinsic effects.

\section{\label{sec:discussion}IV. Discussion}
So far our results have shown a compelling relation between the conducting sample surface, as clearly indicated by the low-$T$ resistivity and changes of the skin depth, and the highly unusual Gd$^{3+}$ ESR line shape in Sm$_{1-x}$Gd$_{x}$B$_6$. 
In analogy to the case of CESR, the Gd$^{3+}$ line shape could be described by a parameter $X/R$ defining a diffusive asymmetry. 
However, the spin probes themselves are not diffusing but their spin excitations do. In order to understand in more detail such relaxation mechanism, we should look into the relevance of the surface effects in the spin-lattice relaxation $T_{1}$.

In metals, the relaxation of the spin probe is reflected in the linewidth, which is proportional to 1/$T_{2}$ \cite{barnes1981theory}. However, here we have a low-$T$ bulk insulator and therefore, any evolution of the relaxation of the system should not necessarily be reflected in $T_{2}$, but in the spin-lattice relaxation $T_{1}$. As such, from the saturation behavior of the ESR intensity $I$ as a function of the microwave power we can indirectly estimate $T_{1}$ (for more details see \cite{poole1971relaxation,supp2020souza}). Figure \ref{Fig4} (f) displays $T_{1}$ as a function of the FIB grid. As a matter of comparison, an estimated $T_{1}$ for Sm$_{1-x}$Gd$_{x}$B$_{6}$ with $x$ = 0.0004, for which the line shape shows no diffusive asymmetry, is also presented. 

The latter comparison reveals a distinct Gd-concentration dependence of $T_{1}$. The minute substitution of Sm by Gd results in a local modification of the Kondo lattice of SmB$_{6}$ and, hence, in a local reduction of the hybridization gap \cite{souza2020metallic,lawrence1996kondo}. This additional disorder gives rise to an extra relaxation channel and reduces $T_{1}$ in an effectively similar way as in the opening of a bottleneck process \cite{barnes1981theory}. With $T_{1}$ being too small due to extra relaxation channels, the line shape is of a localized-type.

Looking now in figure \ref{Fig4} (f) for $x = 0.0002$ at the FIB effects on $T_{1}$ and comparing them with the intensity (reflecting the skin depth) shown in figure \ref{Fig4} (d), we can see that the increase of the skin depth is related to a decrease of the effective $T_{1}$. This finding may have two possible origins. The first one is that the cuts at the surface introduce incoherent (disorder) scattering, which diminishes the spin-lattice relaxation time at the surface.

The other one is related with the importance of relaxations due to the surface as such. Surface relaxations are important whenever there exist strong spin-dependent forces, $i.e.$ spin polarization, during a collision of the spin excitation with the surface \cite{feher1955electron}. F.J. Dyson treats this modification briefly and shows that in the case of thin films or small particles, one expects a much more marked effect than in bulk materials \cite{dyson1955electron}. In other words, the increase of the microwave penetration and the decrease of the mean free path $\ell$, both FIB-induced effects, make the surface relaxation less relevant, which would result in a strongly reduced diffusive asymmetry of the line shape.

The analogies between the ESR line shape characters of conduction electrons and Gd$^{3+}$ in SmB$_6$ are sensible and, hence, constitute a valuable basis for constructing our relaxation mechanism which results in the Gd$^{3+}$ diffusive-like line shape. To this end, it is instructive to first recall how a phonon bottleneck process could be responsible for the coupling between local moments and the surface excitations \cite{lesseux2016unusual,garanin2007towards}. In a phonon bottleneck process, the phonons emitted by a direct process will be reabsorbed by the magnetic ions in the lattice \cite{garanin2007towards}. This is a long-memory effect, meaning that it will result in an effective increase of the spin-lattice relaxation time $T_{1}$. This increase of $T_{1}$ can be interpreted as an increase of the phonon momentum coherence, which we will call long-living phonons \cite{garanin2015angular,nakane2018angular}. In fact, such a spin-phonon process has already been demonstrated experimentally \cite{holanda2018detecting}. In this sense, the increase of the phonon coherence could enhance the probability of an energy transfer to surface excitations prior the relaxation to the thermal bath.

\begin{figure}[!ht]
\includegraphics[width=0.95\columnwidth]{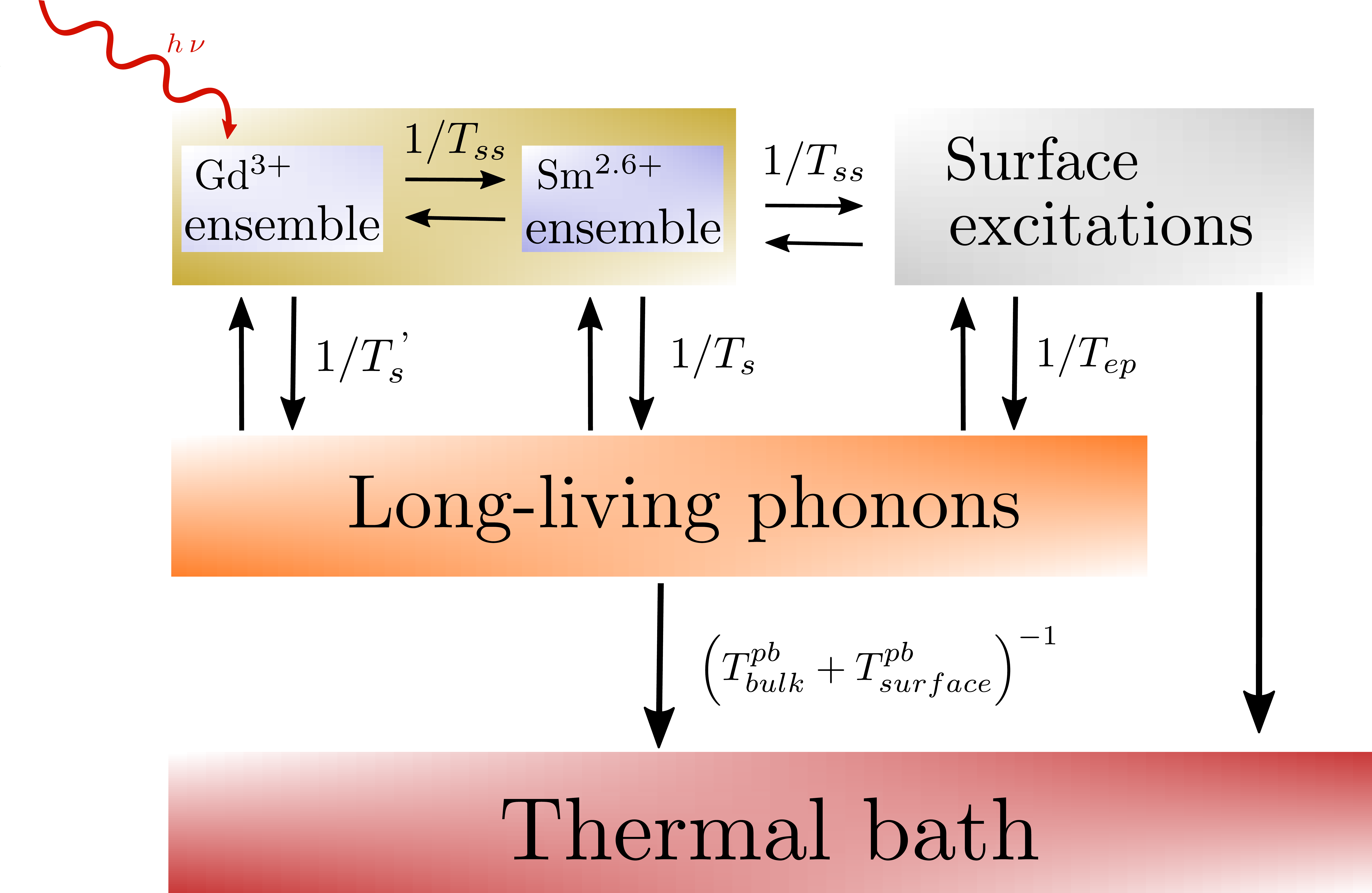}
\caption{Schematic representation of the proposed relaxation mechanism. $1/T_{ss}$ represents a spin-spin relaxation rate, $1/T_{s}$ the spin-phonons relaxation, $1/T_{ep}$ the relaxation through an electron-phonon coupling at the surface and $T^{pb}_{bulk}$ the time scale of the phonons decoherence from the bulk and $T^{pb}_{surface}$ the time scale of the surface phonons decoherence.}
\label{Fig5}
\end{figure}

Figure \ref{Fig5} illustrates the various intermediate processes we propose for the total spin-lattice relaxation $1/T_{1}$. In this scenario we also take into account the relaxation through the surface. The Gd$^{3+}$ relaxation to the phonons is realized through its crystal field \cite{garanin2015angular,nakane2018angular}, which is represented by the rate $1/T'_{s}$, where $T'_{s}$ is the relaxation time from the spin probe to the phonons. Because Gd$^{3+}$ has zero orbital momentum ($L = 0$), such coupling will not be so relevant. Now, we posit that the only reason that it is possible to see the diffusive asymmetry lineshape in Gd$^{3+}$-substituted SmB$_{6}$ is due to the Gd$^{3+}$--Sm$^{2.6+}$ coupling which was discussed in ref. \cite{souza2020metallic}. Because of their large concentration and $L\neq 0$, Sm ions have a much more efficient coupling, $1/T_{s}$ $\gg$ $1/T'_{s}$, with long-living phonons. Such coupling between a concentrated Sm matrix with the phonons results in a phonon-bottleneck process and, concomitantly, an enhancement of $T_{1}$.

Whenever the transferred energy, through the phonons and even Sm$^{2.6+}$ ions, reaches the surface there is an impedance in the thermal exchange with the thermal bath, which is denoted by $(T^{pb}_{bulk} + T^{pb}_{surface})^{-1}$, where $pb$ denotes phonon bottleneck. Due to this impedance it is possible to obtain a coupling of the surface phonons, or even the surface Sm ions, to the surface excitations, which effectively results in a diffusion of the magnetization while the system relaxes to the thermal bath.

One key question is related to the nature of the surface excitations. Indeed, one may argue that they not necessarily need to have a topological, or even an electronic nature. Clearly, without a proper model, which is beyond the scope of this work, we cannot rule out this scenario, but it seems unlikely. There are few points in favor of electronic excitations being the crucial ingredient, probably with a topological character. First our results show a compelling correlation between electrical conductance attributed to surface states and the evolution of the Gd$^{3+}$ ESR line shape. One should also mention the $T$-dependence for $T$ $\leq$ 4 K of the $X/R$ parameter, especially for the sample without FIB-cut trenches, denoted as ``no FIB" in fig. \ref{Fig4} c). It shows, around $T$ $\approx$ 4 K, a reduction of the increase of the $X/R$ parameter, which even appears to saturate. This happens right at the temperature where the surface states start to dominate the resistivity measurements. A similar situation is observed for sample S1, as shown in fig. \ref{Fig3}, where at low $T$ there is a slight change in the increase of $X/R$. Although the analysis of $X/R$ for $T \leq 4 K$ should be taken with caution without a proper model, this is another hint of the connection between surface states and the Gd$^{3+}$ ESR line shape. 

The second point in favor of non-trivial electronic excitations is the comparison of Nd$^{3+}$-substituted YPdBi and YPtBi \cite{lesseux2016unusual,souza2018diffusive}. They show clearly different robustness of the diffusive line shape against external parameters (such as grain size and Nd$^{3+}$-concentration). With both having a similar skin depth, and similarly enhanced $T_{1}$'s, in principle, a similar robustness of the diffusive asymmetry of the line shape should be expected for the two systems if phonons or electronic trivial excitations were responsible for the diffusive asymmetry; however this is not observed experimentally \cite{lesseux2016unusual,souza2018diffusive}.

At this point it is worth to return to the tuning parameters and understand their role in the light of our proposed relaxation mechanism scenario. Regarding the skin depth change, the decrease of the mean free path may diminish the spin-lattice relaxation time at the surface, which is denoted by $T^{pb}_{surface}$. The reduction of the surface relaxation time diminishes the probability of energy transfer to surface excitations. In other words, the role of the surface excitations to the relaxation of the system is suppressed as a function of the FIB cutting grids. Such suppression results in a decrease of the diffusive asymmetry, which is reflected in the $X/R$ parameter, eventually leading to a Gd$^{3+}$ localized-like line shape.

Another explored parameter is the Gd$^{3+}$ concentration. Samples with $x$ = 0.0004 of Gd$^{3+}$ show a plateau in resistivity at low $T$ \cite{souza2020metallic}, however we speculate that in $x$ = 0.0004 the number of Gd$^{3+}$ ions is too high for a relaxation through the surface excitations. Moreover, one should not forget disorder effects. The additional disorder due to the increase of the Gd$^{3+}$ concentration may suppress the spin-lattice relaxation, even at the surface, due to the creation of additional relaxation channels, which introduces incoherent scattering. This is nicely confirmed by the small $T_{1}$ for $x$ = 0.0004 when compared with $x$ = 0.0002 results. In fact, the FIB results, specifically those for small numbers of FIB-cut trenches, show that a tiny amount of disorder already creates effects in the line shape.

Finally, our work suggests that the highly diluted exchange ($x$ = 0.0002) of Sm$^{2.6+}$ ions by Gd$^{3+}$ ions does not affect the local topology around the substituted sites. This is consistent with recent claims that a probe inside of a topological non-trivial matrix could not alter the topological nature of the system \cite{grusdt2016interferometric}. Our results demonstrate the influence of surface excitations in the relaxation of our ESR probe, which leads to an unusual diffusion asymmetry of the ESR line shape. Furthermore, they provide first hints that the unusual ESR line shape is likely related to the topological nature of the system.

\section{\label{sec:conclusion}V. Conclusion}

In summary, we performed electron spin resonance and complementary resistivity measurements in the Kondo insulator Sm$_{1-x}$Gd$_{x}$B$_{6}$ with $x$ = 0.0002 and 0.0004. The Gd$^{3+}$ ESR spectrum at $T$ = 4 K for $x$ = 0.0002 shows a diffusive-like character, which correlates with the temperature evolution of surface states in SmB$_6$. Using a focused ion beam we systematically altered the sample surface and showed the evolution of the Gd$^{3+}$ ESR line shape from a diffusive-like to a localized-like character. Our analysis of the spin-lattice relaxation time $T_{1}$ reveals that the surface impedance opens the possibility of a diffusive asymmetry through non-trivial surface excitations. Further experiments in other systems and a theoretical description would be valuable to gain more insight to how electron spin resonance can be a smoking gun in the study of topological phases of matter.

\begin{acknowledgments}

We thank Dieter Ehlers (University of Augsburg) for implementing fit functions in his Spektrolyst software package. This work was supported by FAPESP\ (SP-Brazil) Grants No 2020/12283-0, 2018/11364-7, 2017/10581-1, National Council of Scientific and Technological Development - CNPq Grants No 309483/2018-2, 141026/2017-0, 442230/2014-1 and 304649/2013-9, CAPES Finance Code 001, FINEP-Brazil and Brazilian Ministry of Science, Technology and Innovation. ZF acknowledges the funding from NSF-1708199. Work at Los Alamos National Laboratory (LANL) was performed under the auspices of the U.S. Department of Energy, Office of Basic Energy Sciences, Division of Materials Science and Engineering.

\end{acknowledgments}

\end{document}